\documentclass{emulateapj}

\shorttitle{Debris Disk Around the Helix Central Star} 

\newcommand{\um}{${\rm \mu m}$~}
\newcommand{\mm}{${\rm \mu m}$}

\slugcomment{{\footnotesize \it Received 2006 December 22; accepted 2007 January 22 by ApJL
    }}

\begin{document}

\title{A Debris Disk around the Central Star of the Helix Nebula?}

\author{
  K. Y. L. Su\altaffilmark{1},
  Y.-H. Chu\altaffilmark{2},
  G. H. Rieke\altaffilmark{1},
  P. J. Huggins\altaffilmark{3},
  R. Gruendl\altaffilmark{2},
  R. Napiwotzki\altaffilmark{4},
  T. Rauch\altaffilmark{5},
  W. B. Latter\altaffilmark{6},
  K. Volk\altaffilmark{7}
}

\altaffiltext{1}{Steward Observatory, University of Arizona, Tucson, AZ; ksu@as.arizona.edu}
\altaffiltext{2}{Department of Astronomy, University of Illinois at
  Urbana-Champaign, Urbana, IL}
\altaffiltext{3}{Physics Department, New York University, New York, NY} 
\altaffiltext{4}{Centre for Astrophysics Research, Science and
  Technology Research Institute, University of Hertfordshire,
  Hatfield, UK} 
\altaffiltext{5}{Institut f\"ur Astronomie und Astrophysik, Universit\"at
  T\"ubingen, T\"ubingen, Germany} 
\altaffiltext{6}{NASA {\it Herschel} Science Center, California
  Institute of Technology, Pasadena, CA} 
\altaffiltext{7}{Gemini Observatory, Hilo, HI}

\begin{abstract}

Excess emission from a point-like source coincident with the central
star of the Helix Nebula is detected with {\it Spitzer} at 8, 24, and
70~\mm. At 24~\mm, the central source is superposed on an extended
diffuse emission region.  While the [\ion{O}{4}] 25.89~\um line
contributes to the diffuse emission, a 10-35~\um spectrum of the
central source shows a strong thermal continuum. 
The excess emission from the star most likely originates from a
dust disk with blackbody temperatures of 90--130 K.  Assuming a simple
optically thin debris disk model, the dust is distributed in a ring
between $\sim$35 and $\sim$150 AU from the central star, possibly
arising from collisions of Kuiper-Belt-like Objects or the break-up of comets
from an Oort-like cloud that have survived from the post-main-sequence
evolution of the central star.

\end{abstract}


\section{Introduction}

The Helix Nebula (NGC 7293), at only 219 pc \citep{harris07}, is one
of the best studied planetary nebulae. {\it Hubble Space Telescope}
({\it HST}) images at optical wavelengths and recent {\it Spitzer
Space Telescope} ({\it Spitzer}) images at 3.6--8~\um have resolved
the Helix Nebula into thousands of ``cometary knots'' immersed in
diffuse gas, producing strikingly beautiful and well-publicized images
\citep{odell04,hora06}. In contrast, {\it Chandra} X-ray observations
of the Helix detected only a point source associated with the central
star, but the X-ray properties of this source are mysterious.

The central star of the Helix Nebula, WD 2226$-$210, is a
$\sim$110,000~K hot white dwarf \citep{napiwotzki99,traulsen05}.  As
expected, its photosphere emits soft X-rays at energies up to
$\sim$0.25 keV; 
however, {\it Chandra} and {\it ROSAT} observations revealed
additional hard X-ray emission near 1 keV
\citep{leahy94,guerrero01}. The luminosity, temporal variations, and
spectral properties of this hard X-ray component \citep{guerrero01}
and the variability of WD 2226$-$210's H$\alpha$ emission line profile
\citep{gruendl01} are consistent with the coronal activity of an M
dwarf.  However, a sensitive {\it HST} search has ruled out the
presence of any companion with a spectral type earlier than M5
\citep{ciardullo99}.

The mystery regarding the central star of the Helix Nebula deepens
with new {\it Spitzer} observations at 24 and 70~\mm, as reported in
this paper.

\section{Observations, Data Reduction and Analysis}  
\label{observation} 

The Helix Nebula was observed with all three {\it Spitzer}
instruments: the Infrared Array Camera \citep[IRAC;][]{fazio04}, the
Infrared Spectrograph \citep[IRS;][]{houck04}, and the Multiband
Imaging Photometer for {\it Spitzer} \citep[MIPS;][]{rieke04}. The
MIPS observations were made in the scan map mode with a medium scan
rate and a half-array offset, resulting in an effective integration of
80 s per pointing and covering an area of 0\fdg5 $\times$ 0\fdg8 in
each of the 24, 70, and 160~\um bands. Three MIPS scan maps were taken
in sequence with $\gtrsim$10 hr separations to avoid confusion by
asteroids.  The IRS high-resolution observations of the central star
and a nearby nebular position were made in sequence in the staring
mode.  The IRAC observations have been described in detail by
\citet{hora06}.

\begin{figure} \figurenum{2}
\label{fig2}
\plotone{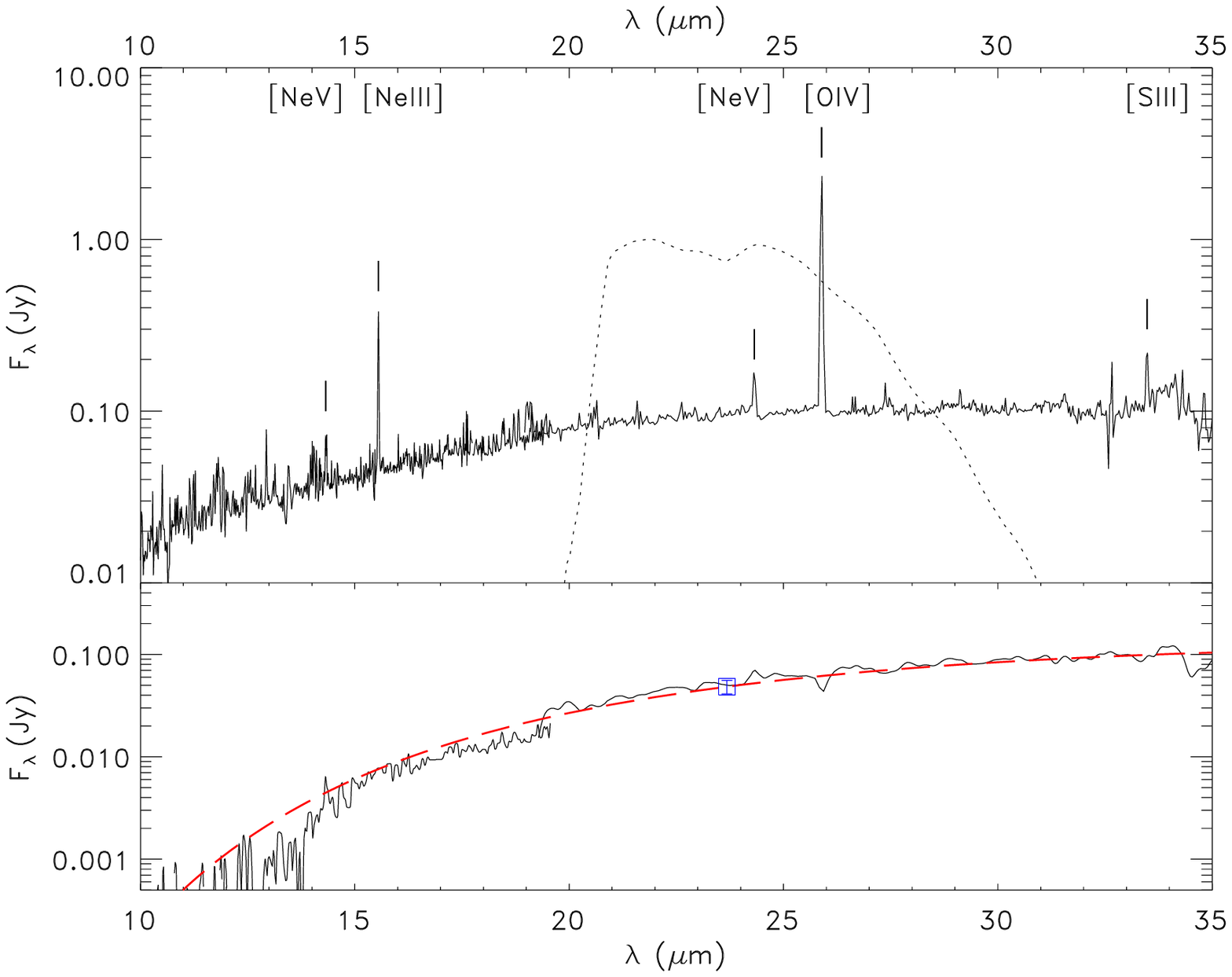}
\caption{{\it Spitzer} {\footnotesize spectrum of the Helix central star. The top panel
  shows the spectrum without sky (nebula) subtraction; therefore, the
  spectrum contains contributions from nebula, zodiacal background and
  the central point source. For comparison, the MIPS 24~\um bandpass
  is shown as a dotted line in the top panel. The bottom panel shows
  the spectrum from the central souce after nebular and sky
  subtraction and Gaussian smoothing (with no scaling between SH and
  LH modules). MIPS 24~\um photometry is shown as a square. A
  blackbody of 102 K is superposed as a long dashed line.}}
\end{figure}

Basic reduction for the MIPS data was carried out using the MIPS Data
Analysis Tool \citep{gordon05}.
%to convert the raw data to calibrated mosaics. 
IRAC mosaics were generated from the {\it Spitzer} Science Center
(SSC) Basic Calibrated Data (BCD) product (ver.~S13.2) using a customized IDL program. The
central 15\farcm4 $\times $12\farcm5 region of the nebula at 3.6, 4.5,
8.0, 24, 70 and 160~\um is displayed in Figure \ref{fig1}.  IRS
high-resolution spectra were extracted from the SSC BCD product
(ver.~S14.0) using the IRS team's Spectral Modeling, Analysis, and
Reduction Tool \citep[SMART;][]{higdon04}. As the spectra on the star
were extracted using a full-slit width mode in both the Short-High
(SH, 9.9--19.6~\mm, 5\arcsec$\times$11\arcsec~slit) and Long-High (LH,
18.7--37.2~\mm, 11\arcsec$\times$22\arcsec~slit) modules without
off-source sky subtraction, the LH spectrum was scaled to match the
continuum level of the SH spectrum. The scaled LH spectrum and the SH
spectrum are plotted together in the top panel of Figure \ref{fig2}.

The IRAC images of the Helix Nebula show a bare point source at the
location of the central star, WD 2226$-$210 (see Fig.~1).  This
central source is detected with flux densities of 374$\pm$19,
241$\pm$24, 171$\pm$26, and 174$\pm$17 $\mu$Jy at the 3.6, 4.5, 5.8,
and 8.0~\mm\ IRAC bands, respectively. The flux densities in the first
three bands are, within error limits, consistent with those expected
from the Rayleigh-Jeans tail of a 110,000 K blackbody model normalized
to the Two Micron All Sky Survey (2MASS) photometry of the star (2MASS
22293854$-$2050136), $J$ = 14.332$\pm$0.024, $H$ = 14.490$\pm$0.039,
and $K_{\rm s}$ = 14.548$\pm$0.079. The flux density at 8~\um is,
however, twice as high as expected.

The MIPS images of the Helix Nebula show very different 
distributions of emission. In the 24~\um band, a pointlike 
source coincident with the central star is superposed on a 
plateau of diffuse emission. In the 70~\um band, no diffuse emission
is seen, and only a central pointlike source is detected. 
In the 160~\um band, neither the central source nor
the bright diffuse emission is detected.

To analyze the central point-like source at 24~\mm, the bright diffuse
background needs to be subtracted. The surface brightness of the
diffuse emission falls off linearly with radius ($r$) from the central
star. Therefore, we use a linear model fit to the surface brightness
profile between $r$ = 31\arcsec\ and 100\arcsec\ to estimate the
diffuse emission superposed on the central point-like source.  After
background subtraction, the central source appears slightly resolved
with a FWHM of $\sim$9\arcsec, 1.5 times that of a true point source,
possibly due to an imperfect subtraction of the background.  Using
aperture photometry and a color correction for a $\sim$100 K blackbody
(see \S \ref{physical_structure} for justification), the
background-subtracted flux density of the central source is
48.4$\pm$7.3 mJy in the 24~\um band.  No variability is detected at
24~\um in our three observations.  Using a similar method, we measure
a flux density of 224$\pm$33 mJy in the 70~\um band and a 3 $\sigma$
upper limit of 237 mJy in the 160~\um band.

\section{Discussion}
\label{discussion}

\subsection{The Diffuse Emission and Point-Source Regions}

The detection of a point-like source in the MIPS 24 and 70~\um bands
is puzzling because the large extent of the ionized nebula is well
resolved with {\it Spitzer}. While the MIPS 24~\um flux from the
unresolved source could originate from emission lines in the bandpass,
the lines would have to originate in high-ionization stages from
abundant elements. Possible lines are the [\ion{Ne}{5}] 24.32~\um line
and [\ion{O}{4}] 25.89~\um line, the latter having been suggested by
\citet{leene87} from the {\it IRAS} observation. The ionization
thresholds needed to produce \ion{O}{4} (54.9 eV) and to excite the
$\lambda$4686 line by the ionization and recombination of \ion{He}{2}
(54.4 eV) are similar. Therefore, we resort to a \ion{He}{2}
$\lambda$4686 image of the Helix Nebula \citep{odell04} to assess the
spatial extent of the [\ion{O}{4}] emission at a higher resolution.
The \ion{He}{2} image of the Helix Nebula shows a central diffuse
emission region similar to that seen at 24~\mm, but not a point-like
central source. From this comparison we conclude that the [\ion{O}{4}]
line contributes to the diffuse emission up to a radius of
$\sim$2\arcmin.

\begin{figure}
\figurenum{3}
\label{fig3}
\plotone{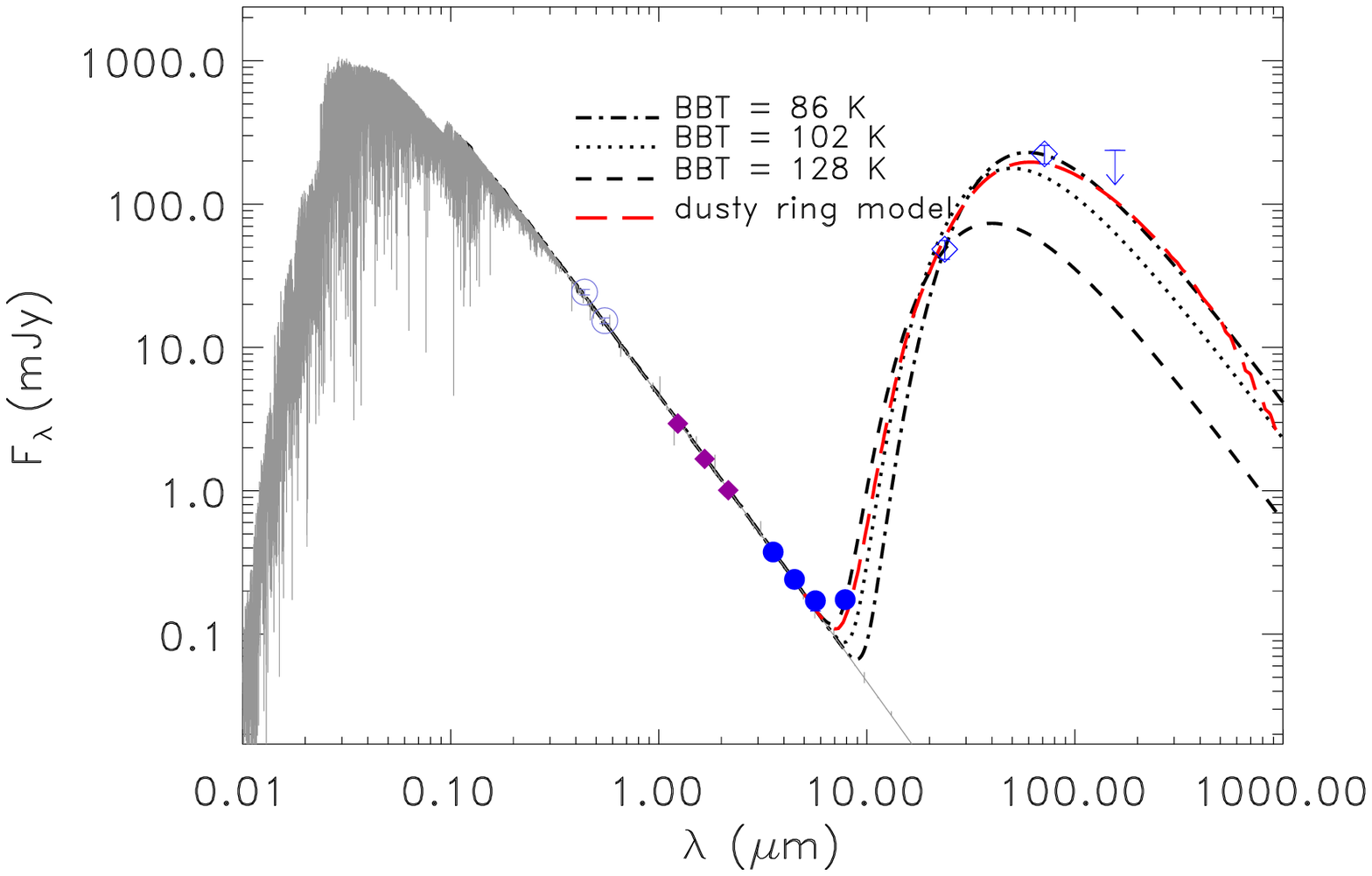}
\caption{ {\footnotesize SED of the central white
dwarf in the Helix Nebula. The photometry for ground-based Johnson $B$
and $V$ \citep{harris07} is displayed as open circles, as filled diamonds for 2MASS,
filled circles for IRAC, and open diamonds for MIPS photometry
with a downward arrow showing the 160~\um 3 $\sigma$ upper limit. 
 The white dwarf photospheric model is shown as a
 grey line. The excess SED is
 consistent with a blackbody with temperatures (BBT) ranging from 86 K (the
 best fit using MIPS 24 and 70~\um points) to 128 K (the best fit
 using IRAC 8~\um and MIPS 24~\mm). An intermediate temperature of 102
 K is also shown as a heavy dotted line for comparison. The SED of a
 dusty ring (35-150 AU) is shown as a (red) long-dashed line.}}
\end{figure}

The absence of the central point-like source in the \ion{He}{2} image
confirms that the [\ion{O}{4}] line does not contribute to the central
peak in the MIPS 24~\um band. 
Another promising
candidate is the [\ion{Ne}{5}] 24.32~\um line. With an excitation
potential of 97 eV, \ion{Ne}{5} can be produced via photoionization by
hard X-rays or thermal collisional ionization at temperatures ($T$)
higher than $ 10^5$ K.  At such high temperatures, the \ion{He}{2}
$\lambda$4686 recombination line may be weakened to below the detection
limit because of its $T^{-1/2}$ dependence. While this might explain
the absence of the central peak in the \ion{He}{2} image, it cannot
explain the central point-like source in the MIPS 70~\um band, as this
band does not contain any nebular lines of abundant elements at high
ionization stages.

Both [\ion{O}{4}] and [\ion{Ne}{5}] lines are seen in our IRS spectrum 
of the central region (see the top panel of Fig.~\ref{fig2}) on top of a continuum
(a combination of dust grains in the ionized nebula, zodiacal
background and the central point source). 
Without a proper subtraction of the nebular and zodiacal contribution, it is
difficult to assess the relative importance of the emission lines and
the continuum from the central star based on this IRS spectrum alone.
The true spectrum of the central
source is revealed after subtraction of the nebular and sky 
contribution (shown in the bottom panel of Fig.~\ref{fig2}). 
The spectrum of the central source will be addressed in more detail 
in another paper.

\subsection{Nature of the Central Infrared Source}

The spectral energy distribution (SED) of the central source is shown in
Figure \ref{fig3}. For the white dwarf, we employ TMAP, the T\"ubingen NLTE Model
Atmosphere Package \citep{werneretal03,rauchdeetjen03}, in its most
recent version for the calculation of plane-parallel, static white
dwarf models 
that consider H--He--C--N--O--Mg--Si--Ca--Ni model atoms. The SED shortward of 
6~\um is consistent with the scaled white dwarf photospheric model for 110,000 K. 
Assuming that the excess infrared emission
originates from a blackbody radiator, its temperature is 90--130 K (see Fig.~\ref{fig3}), 
far too low for a stellar object. 
If the blackbody radiator is heated by the central 
star WD 2226$-$210, of which the effective temperature is 
110,000 K and radius is 0.024 $R_{\sun}$, it would have to be 41--91 AU from the
star.  This radial distance corresponds to 0\farcs19--0\farcs42,
so it cannot be resolved with {\it Spitzer}.
The luminosity of this blackbody radiator, (5.1--11) $\times$10$^{31}$ 
ergs s$^{-1}$, 
requires a total cross section ($\sigma_{tot}$) of
(0.8--8.7) $\times$10$^{27}$ cm$^2$, or 3.8--38 AU$^2$.
Although we do not have any constraint on the geometry of the emission
source, 
only an extended 
dust cloud, most probably a disk can provide such a large emitting
area. 
The extinction toward the star is small based on the 
{\it International Ultraviolet Explorer} data \citep{bohlin82}, consistent with the
219 pc distance. 
The dust disk is likely to be optically thin and  
tilted since the main nebular ring is tilted by 
$\sim$35\arcdeg~with respect to the line of sight \citep{young99}.

Given the evolution of the Helix central star through the asymptotic
giant branch (AGB) and
planetary nebula phases, the presence of a debris disk from the
stellar main-sequence phase is surprising but not entirely
impossible. Recently, dust disks around the white dwarf GD\,362
\citep{becklin05,kilic05} and the white dwarf
G29-38 \citep{zuckerman87,reach05} were reported; these disks are
closer to their central stars and thus have SEDs showing excess
emission starting from the near-IR bands. The Helix central star is 
much hotter and more luminous than these two white
dwarfs (i.e., the Helix central star is much less evolved along its
cooling track than the other two); 
thus it can heat a debris disk at a larger radial
distance. 

\subsection{Physical Properties of the Debris Disk}
\label{physical_structure} 

If the infrared excess around the Helix central star indeed arises
from a debris disk, the dust grains in the disk that dominate the
radiation are likely to be microns to millimeters in size.  
The minimum grain size will be set by blowout by radiation pressure.  
Given the stellar mass of 0.58 $M_{\sun}$ \citep{napiwotzki99,traulsen05} and a
luminosity of 76 $L_{\sun}$ (from the observed photometry and
distance), grains smaller than $\sim$60~\um will be ejected from the
system. For an initial estimate, one can assume that 
the disk consists of grains with sizes (radius of $a$) from
60 to 1000~\um and has a power-law size distribution, $n(a)\propto
a^{-3.5}$, as expected for a collision-dominated disk
\citep{dohnanyi69,tanaka96}. The relatively large grains support using
blackbody-based approximations to characterize the system.  

For a simple estimate, we take the disk dimensions from the preceding
section 
and approximates the grain size distribution by a single size of
100~\mm, characteristic of the grains that dominate the infrared
emission. The dust mass in these grains is (4/3) $\rho_g a
\sigma_{tot}$ where $\rho_g$ is the grain density and $\sigma_{tot}$
is the total area that intercepts the star light. Since the total dust
mass is proportional to $\int n(a)~a^3 da \propto a^{0.5}$, the total
dust mass from the disk is $\sim$0.11 $M_{\earth}$ assuming
$\rho_g$=2.5 g cm$^{-3}$ and $\sigma_{tot}=$6$\times10^{27} $cm$^2$.

This rough estimate can be tested with a more realistic model that
fits the SED, although such models are still subject to degeneracy and
ambiguities.
We assume that the
grains are astronomical silicates \citep{laor93} with a size
distribution as before. The disk is optically thin, is heated only by
WD 2226--210, and has a constant surface number density,
with inner and outer boundaries of $r_{in}$ and $r_{out}$.  
The observed SED is fitted by a ring extending from 35 to 150 AU with
a total dust mass of 0.13$M_{\earth}$($\pm$50\%), as shown in Figure
\ref{fig3}. 

The exact inner radius of the disk is subject to the grain properties
used in the model; however, the inner radius cannot be too close to
the star because no observable excess is found at wavelengths shorter
than 8~\mm. An inner hole is required by the SED modeling. The outer
boundary of the disk is constrained in our model because a disk
with a much larger outer radius would have higher 160~\um flux,
contradicting the observed upper limit. Therefore, the dust around the
Helix central star is in a form of ring between $\sim$35 and $\sim$150
AU. This configuration is remarkably similar to the planetary debris
disks seen around main-sequence stars.

\subsection{Origin of the Debris Disk}

What is the origin of the dust {\it Spitzer} detects around the Helix
central star? It has been established that any planet closer
than $\sim$1 AU will be engulfed by an expanding red giant
\citep{siess99a,siess99b}, while planets outside $\sim$5 AU
from the Sun will survive post-main-sequence evolution
\citep{sackmann93,debes02}, and the orbits of surviving planets and
most of the Kuiper Belt objects (KBOs) and Oort Cloud comets will
expand adiabatically and remain bound to the solar system
\citep{duncan98}. 
The re-stabilized KBOs and Oort Cloud can later
become the source of objects that go into the inner part of the
system, either plunging into the white dwarf or breaking up due to tidal destruction,
and they can populate the inner system with dust. A similar scenario has
been proposed to explain the photospheric metal contamination in DAZ
white dwarfs \citep{stern90, parriott98,jura06} and may even produce the
observed X-ray emission at 1 keV. Furthermore, \citet{debes02}
suggest that if the surviving planet system becomes unstable after the
AGB phase, the entire system becomes dynamically young; i.e., frequent
collisions and encounters can occur among surviving planets and
comets, leading to a period of enhanced ``late bombardment''.  
It is likely that the kilometer-size objects surviving around
the Helix central star are experiencing similar perturbations and
producing fine dust in the system. % that {\it Spitzer} detected. 
The Poynting-Robertson (P-R) lifetime is $3-50\times10^{6}$ yr for
$\sim$100~\um grains located at 35 and 150 AU, while the collisional
lifetime is $0.1-1\times10^{6}$ yr. Note that these timescales are
much longer than the kinematic age of the nebula (2.8$\times10^4$ yr; 
\citealt{napiwotzki99}) and a typical timescale in the post-AGB
evolution ($\sim4\times10^4$ yr for a 0.565 $M_{\sun}$ white dwarf; 
\citealt{schoenberner83}). Assuming an average dust lifetime of
$1\times10^{6}$ yr, the dust production rate is $2.5\times10^{13}$
g~s$^{-1}$. This rate is only a lower limit since the dust production
is not in equilibrium, but it is $\sim$$10^{5}$ 
times the dust production rate observed from comet Hale-Bopp, consistent
with the hypothesis of an enhanced late bombardment in the
post-main-sequence evolution.  

The total observed dust mass, $\sim$0.13 $M_{\earth}$, is equivalent
to 20 asteroidal objects with a radius of $\sim$1200 km. While 
the dust cloud around the white dwarf G29--38 may arise from 
tidal disruption of Pluto-sized objects \citep{jura03}, this cannot
be the case for the disk in WD 2226--210 because the tidal 
disruption zone is much closer to the central star, $\sim$0.003 AU 
(\citealt{jura03}, eq.[4]). It is thus unlikely that the debris disk is 
produced by tidal breaking of large asteroidal objects, but rather 
by enhanced collisions among the surviving Kuiper Belt-like and/or 
cometary objects.

The inner radius of a P-R drag-dominated disk, in general, should be
very close to the star, up to the dust sublimation radius ($\sim$0.1
AU in the case of the Helix central star)
if there is no planet-size or sub-stellar object inside the disk to
stop the dust grains from spiraling inward. The inner hole ($r$$
\lesssim$35 AU) derived from our SED model is likely caused by the
dynamical perturbation of unseen planets or sub-stellar objects inside
the hole, which can trap the migrating dust or scatter the dust
outward \citep{meyer06}. Alternatively, the inner hole may also be
explained by ice sublimation if the grains are icy \citep{jura98},
a process that will occur at a dust temperature of $\sim$150~K, 
corresponding to $\sim$30 AU.

If the Helix central star had a close binary in the main-sequence
phase, a binary merger event could produce an accretion disk around
the remaining white dwarf. The physical structure of the disk would be
highly dependent on the properties of the original binary system;
however, the fact that there is no near-infrared excess suggests that
the inner part of the optically thick accretion disk would have to
dissipate in a short period of time after the binary merged. It is
hard to explain why the dust is so far away from the white dwarf if it
was the remnant disk due to close binary evolution. Alternatively if
the companion was in an intermediate orbit and avoided spiraling into
the primary during the evolution, the binary configuration is
efficient at trapping the dust from the mass loss of the primary to
form a stable circumbinary disk in the post-AGB phase
\citep{vanwinckel03,deruyter06}, inducing the presence of the dusty
disk we detected. However, any of the micron-size dust generated in
the AGB phase should be blown out by the strong radiation in the early
white dwarf evolution. The fact that we do see dust around the Helix central
star favors the debris disk scenario. Future high spatial resolution
infrared imaging is needed to further constrain the geometry of the
emitting region.

\acknowledgments

This work is based on observations made with the {\it Spitzer Space
Telescope}, which is operated by the Jet Propulsion Laboratory,
California Institute of Technology, under NASA contract 1407. 
Code for calculating optical properties of grains at large size
parameters kindly provided by Viktor Zubko and Karl Misselt. 
Support for this work was provided by NASA through contract 1255094, issued by
JPL/Caltech. T.R\@. was supported by the BMBF/DESY under grant 05\,AC6VTB.

\begin{figure*}
\figurenum{1}
\label{fig1}
\plotone{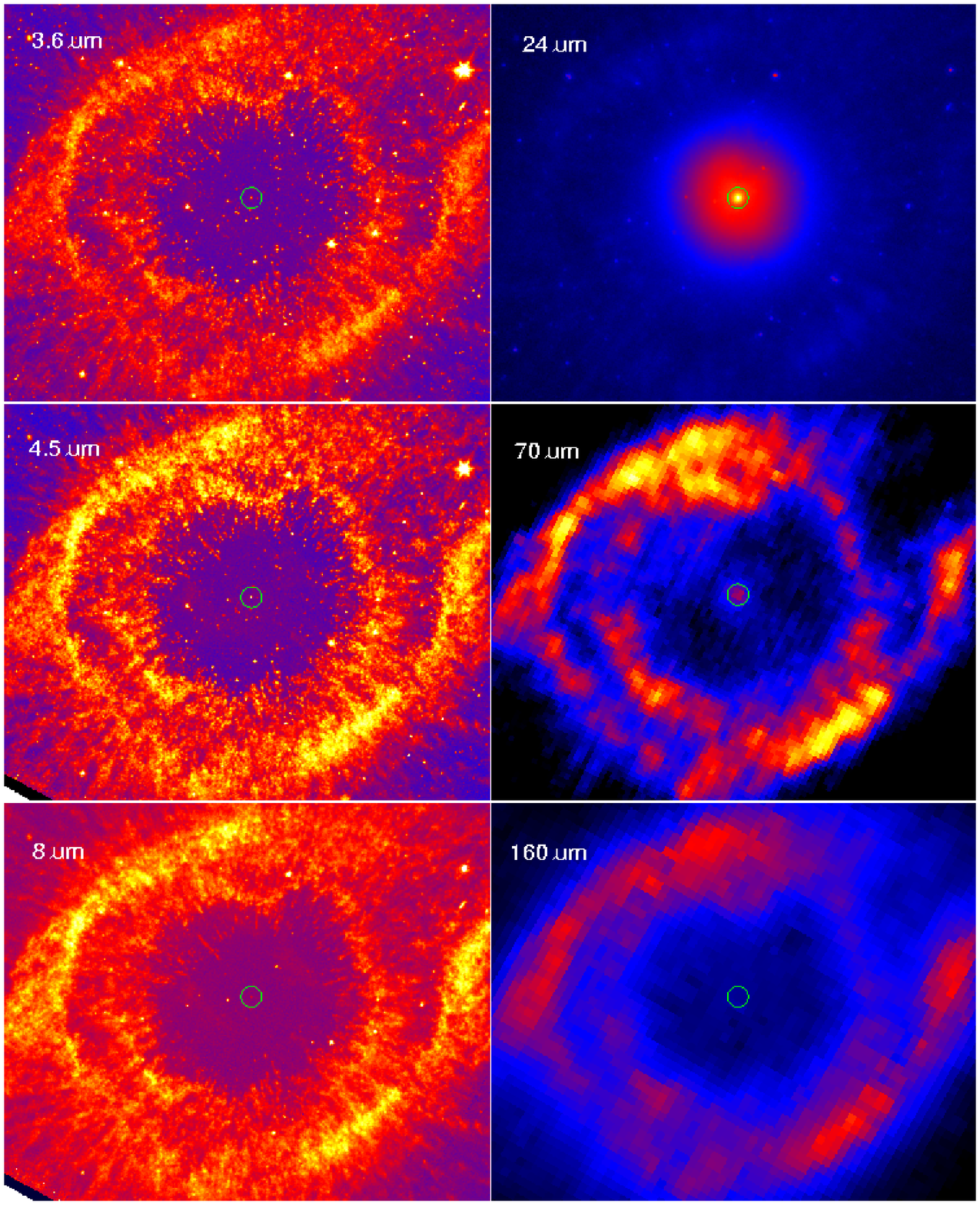} 
\caption{{\it Spitzer} images of the Helix Nebula. Each panel is 
displayed in a false-color logarithmic scale with brightness and
contrast adjusted for best presentation and oriented as north up and east
toward the left with a field of view of 15\farcm4$\times$12\farcm5. 
}
\end{figure*}


\begin{thebibliography}{}

{\footnotesize 
\bibitem[Becklin et al.(2005)]{becklin05} Becklin, E.~E., Farihi, 
J., Jura, M., Song, I., Weinberger, A.~J., \& Zuckerman, B.\ 2005, \apjl, 
632, L119 

\bibitem[Bohlin et al.(1982)]{bohlin82} Bohlin, R.~C., 
Harrington, J.~P., \& Stecher, T.~P.\ 1982, \apj, 252, 635 

\bibitem[Ciardullo et al.(1999)]{ciardullo99} Ciardullo, R., Bond, 
H.~E., Sipior, M.~S., Fullton, L.~K., Zhang, C.-Y., \& Schaefer, K.~G.\ 
1999, \aj, 118, 488

\bibitem[Debes \& Sigurdsson(2002)]{debes02} Debes, J.~H., \& 
Sigurdsson, S.\ 2002, \apj, 572, 556 

\bibitem[de Ruyter et al.(2006)]{deruyter06} de Ruyter, S., van 
Winckel, H., Maas, T., Lloyd Evans, T., Waters, L.~B.~F.~M., \& Dejonghe, 
H.\ 2006, \aap, 448, 641 

\bibitem[Dohnanyi(1969)]{dohnanyi69} Dohnanyi, J.~W.\ 1969, \jgr, 74, 2531 

\bibitem[Duncan \& Lissauer(1998)]{duncan98} Duncan, M.~J., \& 
Lissauer, J.~J.\ 1998, Icarus, 134, 303 

\bibitem[Fazio et al.(2004)]{fazio04} Fazio, G.~G., et al.\ 
2004, \apjs, 154, 10 

\bibitem[Gordon et al.(2005)]{gordon05} Gordon, K.~D., et al.\ 2005, \pasp, 117, 503 

\bibitem[Gruendl et al.(2001)]{gruendl01} Gruendl, R.~A., Chu, 
Y.-H., O'Dwyer, I.~J., \& Guerrero, M.~A.\ 2001, \aj, 122, 308 

\bibitem[Guerrero et al.(2001)]{guerrero01} Guerrero, M.~A., Chu, 
Y.-H., Gruendl, R.~A., Williams, R.~M., \& Kaler, J.~B.\ 2001, \apjl, 553, 
L55 


\bibitem[Harris et al.(2007)]{harris07} Harris, H.~C., et al.\ 
2007, \aj, 133, 631 

\bibitem[Higdon et al.(2004)]{higdon04} Higdon, S.~J.~U., et 
al.\ 2004, \pasp, 116, 975

\bibitem[Hora et al.(2006)]{hora06} Hora, J.~L., Latter, W.~B., 
Smith, H.~A., \& Marengo, M.\ 2006, \apj, 652, 426 


\bibitem[Houck et al.(2004)]{houck04} Houck, J.~R., et al.\ 
2004, \apjs, 154, 18 

\bibitem[Jura(2003)]{jura03} Jura, M.\ 2003, \apjl, 584, L91 

\bibitem[Jura(2006)]{jura06} Jura, M.\ 2006, \apj, 653, 613 

\bibitem[Jura et al.(1998)]{jura98} Jura, M., Malkan, M., 
White, R., Telesco, C., Pina, R., \& Fisher, R.~S.\ 1998, \apj, 505, 897 

\bibitem[Kilic et al.(2005)]{kilic05} Kilic, M., von Hippel, 
T., Leggett, S.~K., \& Winget, D.~E.\ 2005, \apjl, 632, L115 

\bibitem[Laor \& Draine(1993)]{laor93} Laor, A.~\& Draine, 
B.~T.\ 1993, \apj, 402, 441 

\bibitem[Leahy et al.(1994)]{leahy94} Leahy, D.~A., Zhang, 
C.~Y., \& Kwok, S.\ 1994, \apj, 422, 205 

\bibitem[Leene \& Pottasch(1987)]{leene87} Leene, A., \& 
Pottasch, S.~R.\ 1987, \aap, 173, 145 


\bibitem[Meyer et al.(2006)]{meyer06} Meyer, M.~R., Backman, 
D.~E., Weinberger, A.~J., \& Wyatt, M.~C.\ 2007, in Protostars and Planets V, 
(Tucson: Univ. Arizona Press), 573 

\bibitem[Napiwotzki(1999)]{napiwotzki99} Napiwotzki, R.\ 1999, \aap, 
350, 101 

\bibitem[O'Dell et al.(2004)]{odell04} O'Dell, C.~R., 
McCullough, P.~R., \& Meixner, M.\ 2004, \aj, 128, 2339

\bibitem[Parriott \& Alcock(1998)]{parriott98} Parriott, J., \& 
Alcock, C.\ 1998, \apj, 501, 357 

\bibitem[Rauch \& Deetjen(2003)]{rauchdeetjen03} Rauch, T., \& Deetjen,
J.L. 2003, in ASP Conf. Ser. 288, Workshop on Stellar Atmosphere Modeling,
eds. I. Hubeny, D. Mihalas, \& K. Werner, 103 

\bibitem[Reach et al.(2005)]{reach05} Reach, W.~T., Kuchner, 
M.~J., von Hippel, T., Burrows, A., Mullally, F., Kilic, M., \& Winget, 
D.~E.\ 2005, \apjl, 635, L161 

\bibitem[Rieke et al.(2004)]{rieke04} Rieke, G.~H., et al.\ 
2004, \apjs, 154, 25 


\bibitem[Sackmann et al.(1993)]{sackmann93} Sackmann, I.-J.,  
Boothroyd, A.~I., \& Kraemer, K.~E.\ 1993, \apj, 418, 457 

\bibitem[Sch\"oenberner(1983)]{schoenberner83} Sch\"oenberner, D.\ 1983, 
\apj, 272, 708 

\bibitem[Siess \& Livio(1999a)]{siess99a} Siess, L., \& Livio, 
M.\ 1999a, \mnras, 304, 925 

\bibitem[Siess \& Livio(1999b)]{siess99b} Siess, L., \& Livio, 
M.\ 1999b, \mnras, 308, 1133 

\bibitem[Stern et al.(1990)]{stern90} Stern, S.~A., Shull, 
J.~M., \& Brandt, J.~C.\ 1990, \nat, 345, 305 

\bibitem[Tanaka et al.(1996)]{tanaka96} Tanaka, H., Inaba, S., 
\& Nakazawa, K.\ 1996, Icarus, 123, 450


\bibitem[Traulsen et al.(2005)]{traulsen05} Traulsen, I., 
Hoffmann, A.~I.~D., Rauch, T., Werner, K., Dreizler, S., \& Kruk, J.~W.\ 
2005, in ASP Conf.~Ser.~334, 325 

\bibitem[van Winckel(2003)]{vanwinckel03} van Winckel, H.\ 2003, 
\araa, 41, 391 

\bibitem[Werner et al.(2003)]{werneretal03} Werner, K., Dreizler, S.,
Deetjen, J.L., Nagel, T., Rauch, T., \& Schuh, S.L. 2003, in ASP
Conf. Ser. 288, Workshop on Stellar Atmosphere Modeling, (San Francisco: ASP), 31 

\bibitem[Young et al.(1999)]{young99} Young, K., Cox, P., 
Huggins, P.~J., Forveille, T., \& Bachiller, R.\ 1999, \apj, 522, 387 

\bibitem[Zuckerman \& Becklin(1987)]{zuckerman87} Zuckerman, B., \& 
Becklin, E.~E.\ 1987, \nat, 330, 138 

}
\end{thebibliography}
\end{document}